\documentclass[a4paper,orivec,runningheads]{llncs}
\usepackage{makeidx}  
\usepackage{graphicx}
\usepackage{subfigure}
\usepackage{amsmath}
\usepackage{amssymb}
\usepackage{verbatim}
\usepackage{algorithm}
\usepackage{algorithmic}
\usepackage{booktabs}
\usepackage{multirow}
\usepackage{pbox} 

\begin{document}

\pagenumbering{arabic}

%

%
%
\title{Uncertainty-Guided Efficient Interactive Refinement of Fetal Brain Segmentation from Stacks of MRI Slices }

\titlerunning{Uncertainty-Guided Interactive Refinement of Fetal Brain Segmentation}  
%

\author{Guotai Wang\inst{1,2} 
\and Michael Aertsen\inst{3}
\and Jan Deprest \inst{4, 5}
\and S\'ebastien Ourselin\inst{2}
\and Tom Vercauteren\inst{2}
\and Shaoting Zhang \inst{1, 6}}

\authorrunning{G. Wang, et al.} 

\institute{
$^1$ School of Mechanical and Electrical Engineering, University of Electronic Science and Technology of China, Chengdu, China \\
$^2$School of Biomedical Engineering \& Imaging Sciences, King's College London, London, UK\\
$^3$Department of Radiology, University Hospitals KU Leuven, Leuven, Belgium \\
$^4$ Department of Obstetrics and Gynaecology, University Hospitals KU Leuven, Leuven, Belgium \\
$^5$Institute for Women’s Health, University College London, London, UK \\
$^6$ SenseTime Research, Shanghai, China \\
\email{guotai.wang@uestc.edu.cn}\\
}
%
\maketitle              

\begin{abstract}

Segmentation of the fetal brain from stacks of motion-corrupted fetal MRI slices is important for motion correction and high-resolution volume reconstruction. Although Convolutional Neural Networks (CNNs) have been widely used for automatic segmentation of the fetal brain, their results may still benefit from interactive refinement for challenging slices. To improve the efficiency of interactive refinement process, we propose an Uncertainty-Guided Interactive Refinement (UGIR) framework. We first propose a grouped convolution-based CNN to obtain multiple automatic segmentation predictions with uncertainty estimation in a single forward pass, then guide the user to provide interactions only in a subset of slices with the highest uncertainty. A novel interactive level set method is also proposed to obtain a refined result given the initial segmentation and user interactions. Experimental results show that: (1) our proposed CNN obtains uncertainty estimation in real time which correlates well with mis-segmentations, (2) the proposed interactive level set is effective and efficient for refinement, (3) UGIR obtains accurate refinement results with around 30\% improvement of efficiency by using uncertainty to guide user interactions. Our code is available online\footnote{\url{https://github.com/HiLab-git/UGIR}}.

\end{abstract}
\section{Introduction}

Due to the good soft tissue contrast, fetal Magnetic Resonance Imaging (MRI) is an important tool for diagnosis of abnormalities of the fetal brain during pregnancy~\cite{Keraudren2014}. However, MRI is susceptible to motion of fetuses during scanning. To mitigate this problem, fast imaging techniques are often used to obtain stacks of 2D slices that have good in-plane image quality but suffer from large inter-slice motion and low 3D resolution. Segmentation of the fetal brain from stacks of fetal MRI slices plays a critical role for correcting the inter-slice motion and reconstructing a high-resolution 3D volume for fetal brain studies~\cite{Ebner2020,Keraudren2014,Salehi2017isbi}.

Despite the fact that deep learning with Convolutional Neural Networks (CNNs) has obtained state-of-the-art performance for automatic fetal brain segmentation from fetal MRI~\cite{Ebner2020,Salehi2017isbi}, it is still difficult for these automatic segmentation methods to obtain accurate results when dealing with images with motion artifacts, abnormal appearances due to pathologies and some challenging local regions~\cite{Salehi2017isbi}. To address this problem, an efficient interactive method to refine the automatic segmentation result is highly desirable in practice, which makes the segmentation more accurate and robust to be clinically useful~\cite{Wang2018,Wang2018c}. 

In the literature, some recent works~\cite{Wang2018c,Zhou2019} use a second CNN that takes the initial automatic segmentation result and additional user interactions as input to obtain a refined result. In~\cite{Wang2018}, image-specific fine-tuning and Graph Cut~\cite{Boykov2001} were used for interactive refinement. Though these methods achieved higher accuracy and efficiency than traditional interactive segmentation methods~\cite{Zhao2013}, when used to segment a volumetric data, they require the user to carefully check the initial segmentation in 2D views slice-by-slice and manually identify mis-segmented regions to give interactions for refinement. For stacks of fetal MRI slices,  automatic CNNs could obtain accurate initial segmentation for most slices~\cite{Ebner2020,Salehi2017isbi}, manually identifying mis-segmented regions may be unnecessary for accurately segmented slices while sometimes difficult for challenging slices. Therefore, the efficiency of such methods is limited. 

To improve the efficiency for user interactions, leveraging the uncertainty information of the initial segmentation has been shown to be a promising method~\cite{Top2011} as it can automatically identify potential mis-segmentations and guide the user to give interactions only in some uncertain regions. Despite the availability of several reliable uncertainty estimation methods for CNN-based segmentation, such as Monte Carlo (MC) dropout~\cite{Gal2016}, model ensemble~\cite{Lakshminarayanan2017,Jungo2019a} and test-time augmentation~\cite{Wang2019neurocomp}, they require multiple forward passes at inference time and cannot provide real-time uncertainty estimation for guiding user interactions. Alternatively, a Bayesian Network has been proposed for fast uncertainty estimation with a single forward pass~\cite{Jena2019}. However, their utility for guiding interactive refinement has not been investigated.    

In this paper, we propose a novel uncertainty-guided framework for interactive refinement of automatic segmentation obtained by CNNs and apply it to fetal brain segmentation from fetal MRI. The contribution is three-fold. First, to obtain real-time uncertainty estimation, we propose a novel method using Grouped Convolution (GC)-based CNNs. It obtains multiple predictions simultaneously with a single model and gives uncertainty estimation in a single forward pass, which is more suitable in the scenario of interactive refinement. Second, we propose to guide the user to give interactions more efficiently during refinement according to the uncertainty information. Thirdly, we propose a novel interactive level set method that incorporates the initial segmentation and user interactions in a uniformed framework to obtain accurate refined results efficiently. The superiority of our framework over existing methods was validated in the task of fetal brain segmentation from stacks of motion-corrupted fetal MRI slices.   

\section{Methods}
\begin{figure*}[t]
	\centering 
	\includegraphics[width=1.0\textwidth]{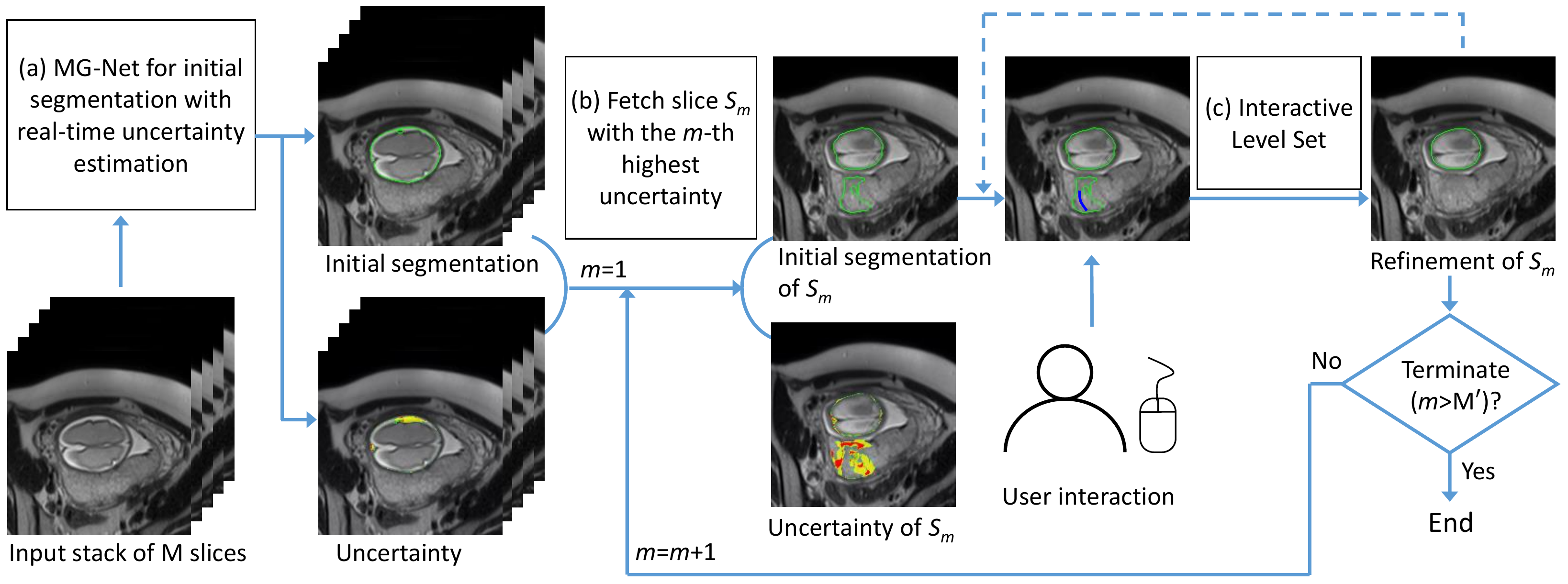}
	\caption{
		The proposed Uncertainty-Guided Interactive Refinement (UGIR) framework for fetal brain segmentation. For an input stack with $M$ slices, the user only needs to give interactions in a subset of $M' (M' < M)$ slices with the highest uncertainty. }
	\label{fig:framework}
\end{figure*}
Our proposed Uncertainty-Guided Interactive Refinement (UGIR) framework for efficient interactive  fetal brain segmentation is shown in Fig.~\ref{fig:framework}. First, a novel CNN based on convolution in Multiple Groups (MG-Net) simultaneously obtains the initial segmentation and uncertainty estimation of a stack of MRI slices in real time. Let $S_m$ ($m$ = 1, 2, ...) denote the slice with the $m$-th highest uncertainty. Our framework automatically and iteratively fetches slice $S_m$ as suggestion for user interactions. After the user gives interactions in $S_m$, a novel interactive level set method obtains the refined result of $S_m$. The iterative refinement is finished when no further slice is suggested by the framework. 
\subsubsection{Simultaneous Initial Segmentation and Uncertainty Estimation.} 
\begin{figure*}[t]
	\centering 
	\includegraphics[width=1.0\textwidth]{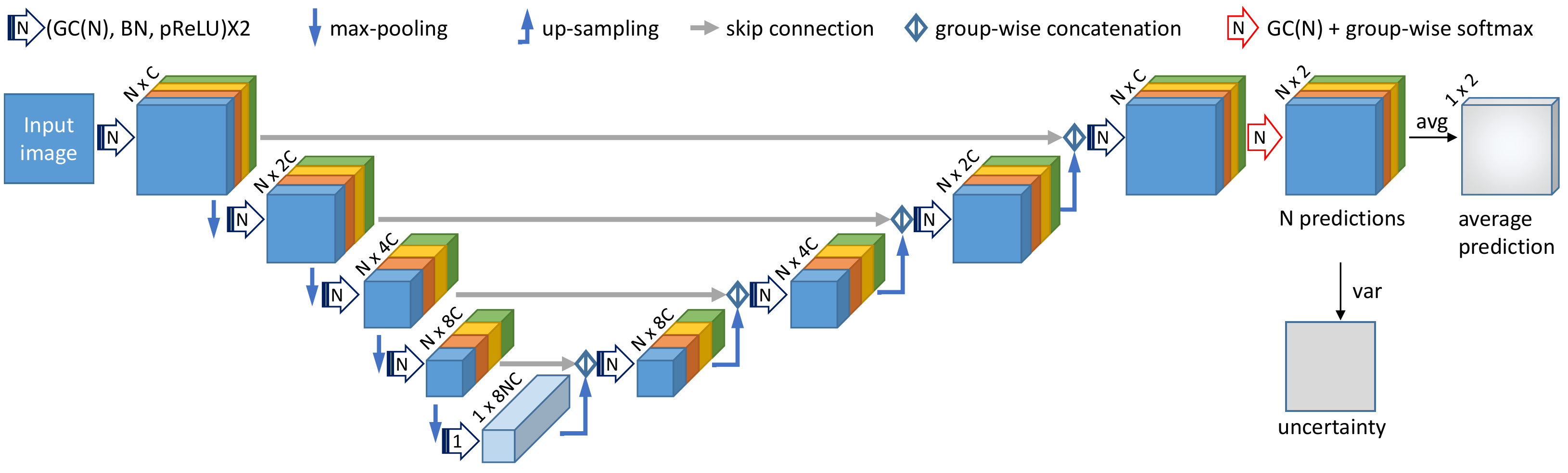}
	\caption{
		The proposed MG-Net that obtains $N$ (e.g., 4) predictions corresponding to $N$ groups of features. The numbers over each feature map such as $N\times C$ are group number $\times$ channel number of each group.  GC(N) denotes convolution with $N$ groups.}
	\label{fig:network}
\end{figure*}
We propose MG-Net to obtain simultaneous initial segmentation and uncertainty estimation in real time with a single network and a single forward pass for inference, which is more efficient for interactive segmentation than typical uncertainty estimation methods including MC dropout~\cite{Gal2016} and model ensemble~\cite{Jungo2019a,Lakshminarayanan2017}. As shown in Fig.~\ref{fig:network}, we modify U-Net~\cite{Ronneberger2015} by using Grouped Convolution~\cite{Krizhevsky2012}. 

An $N$-grouped convolution layer splits the input feature map along the channel dimension into $N$ groups  each with $C_i$  channels. For each group, it uses a convolution kernel of shape $C_o\times C_i \times h \times w$ respectively, where $h \times w$ is the spatial size of the kernel. Therefore, $N$ independent feature maps are obtained each with $C_o$ channels. They are concatenated into a single one with $N\times~C_o$ channels. Note that there is no correlation between different groups in the output feature map of an $N$-grouped convolution layer. Similarly, we implement up-sampling layers by transposed $N$-grouped convolutions, and extend standard channel concatenation to group-wise concatenation to keep the $N$ groups independent of each other. Let $F_1$ and $F_2$ represent two feature maps each with $N$ groups. We first concatenate the $n$-th group of $F_1$ with the $n$-th group of $F_2$, and denote the result as $\hat{F}_n$. Then $\hat{F}_1$, $\hat{F}_2$, ..., and $\hat{F}_N$ are concatenated as the group-wise concatenation of $F_1$ and $F_2$. The group-wise concatenation is used for skip connection between feature maps in the encoder and their counterparts in the decoder, as shown in Fig.~\ref{fig:network}. At the last layer of the decoder, we apply softmax to each feature group respectively (i.e., group-wise softmax) to obtain $N$ probability predictions. Therefore, MG-Net can be seen as an ensemble of $N$ parallel sub-networks, and they are randomly initialized and trained with dropout to obtain diversity.  At the lowest resolution level of  MG-Net, we set group number to one to allow communication of these $N$ sub-networks for better performance. 
 
For training, we apply the same segmentation loss function (i.e., Dice loss~\cite{Milletari2016}) to the $N$ predictions respectively and average their loss values for back-propagation. At inference time, we take the pixel-wise mean value and variance of the  $N$ probability predictions as the final segmentation probability map and pixel-level uncertainty estimation, respectively.  The mean  foreground probability map is thresholded by 0.5 to obtain a binary segmentation mask.

\subsubsection{Uncertainty-Guided User Interactions.} 
 For a  stack of $M$ slices, the CNN is able to obtain accurate segmentation for most slices and only a few slices may require manual refinement~\cite{Salehi2017isbi}. To avoid unnecessary and time-consuming manual search for mis-segmentations of each slice, we ask the user to check and refine only $M'$ ($M' < M$) slices with the highest slice-level uncertainty. 
For a slice $S$, its binary segmentation result $Y$ and pixel-level uncertainty map $U$, a naive slice-level uncertainty can be defined as $\nu^* = \sum_{\mathbf{x}}U_\mathbf{x}$ where $U_\mathbf{x}$ is the uncertainty of pixel $\mathbf{x}$. However, this may lead a slice with a small target to be neglected as it often has a low value of  $\nu^*$ due to a small uncertain region. To address this problem, we alternatively define the slice-level uncertainty as  $\nu = (\sum_{\mathbf{x}}U_\mathbf{x}) / (\sum_{\mathbf{x}}Y_\mathbf{x} + \zeta)$, which is normalized by the segmented region size. $\zeta$ is a small number for numerical stability. We iteratively fetch the slice $S_m$ with the $m$-th highest $\nu$ to ask for interactive refinement, where $m$ = 1, 2, ..., and the iteration terminates when  $m > M'$, as shown in Fig.~\ref{fig:framework}. $M'$ is a predefined number, such as 60\% of $M$ according to the performance of the CNN on the validation set. We also use an early termination strategy when refinement is not needed for three consecutive fetched slices.

\subsubsection{Interaction-based Level Set for Fast Refinement.} For interactive refinement, we use a Distance Regularized Level Set Evolution (DRLSE)~\cite{ChunmingLi2010} due to its efficiency and extend it with an interaction-constraint term, which is named as I-DRLSE. Let $\phi$ denote the level set function that is initialized as the signed distance transform of the initial segmentation result. We define an energy function as $E(\phi) = \alpha E_{r} + \beta E_{u} + \lambda E_{l} +  \mu E_{d}$, where $\alpha$, $\beta$, $\lambda$, $\mu$ are weighting parameters. $E_{r}$, $E_{u}$, $E_{l}$ and  $E_{d}$ are the region, user-interaction, length and distance regularization terms, respectively.  $E_{l} = \int_{\Omega}
\delta_{\epsilon}(\phi(\mathbf{x})) |\nabla\phi(\mathbf{x})|
d\mathbf{x}$ and $E_{d} = \int_{\Omega} p(|\nabla\phi(\mathbf{x})|) d\mathbf{x}$, where $\delta_{\epsilon}$ and $p()$ are the smoothed Dirac delta function and double-well potential function as  in~\cite{ChunmingLi2010} respectively. As the target region in the fetal MRI image has an inhomogeneous appearance that brings  challenges to standard intensity-based level set methods, we define the region term based on the foreground probability map $P$ obtained by MG-Net instead of the original image:
\begin{align}
E_{r} = \int_{\Omega}\Big(
|P - c_1|^{2}H_{\epsilon}(\phi(\mathbf{x})) +
|P - c_2|^{2}(H_{\epsilon}(-\phi(\mathbf{x})))
\Big)d\mathbf{x}
\end{align}
where $H_{\epsilon}$ is the smoothed Heaviside function as in~\cite{ChunmingLi2010}. $c_1$ and $c_2$ are the average foreground probability inside and outside the current level set contour, respectively. Our proposed user-interaction term is:
\begin{align}
E_{u} = -\int_{\Omega}\Big(
H_{\epsilon}(\phi(\mathbf{x}))\text{log}(\eta(\mathbf{x}))
+ H_{\epsilon}(-\phi(\mathbf{x}))\text{log}(1-\eta(\mathbf{x}))
\Big)d\mathbf{x}
\end{align}
where $\eta(\mathbf{x})$ is user-interaction-derived likelihood of  pixel $\mathbf{x}$ being the foreground. Let $\mathcal{F}$ and $\mathcal{B}$ represent the set of pixels specified as the foreground and background by the user interactions, respectively. Inspired by~\cite{Wang2018c}, we use $g_\mathbf{x}^\mathcal{F}$ to represent the geodesic distance between $\mathbf{x}$ and $\mathcal{F}$, and define
$\eta(\mathbf{x}) = e^{-G_\mathbf{x}^\mathcal{F}} / ( e^{-G_\mathbf{x}^\mathcal{F}} + e^{-G_\mathbf{x}^\mathcal{B}})$, where $G_\mathbf{x}^\mathcal{F} = \min(g_\mathbf{x}^\mathcal{F}, D)$
and $G_\mathbf{x}^\mathcal{B} = \min(g_\mathbf{x}^\mathcal{B}, D)$ and $D$ is a threshold value to ensure that only a local region is affected by the interactions. We set $g_\mathbf{x}^\mathcal{F}$ or $g_\mathbf{x}^\mathcal{B}$ as $D$ when  $\mathcal{F}$ or $\mathcal{B}$ is empty. $E_{u}$ is infinite if the segmentation results conflict with the user interactions.

\section{Experiments and Results}
\subsubsection{Data and Implementation.} MRI scans of 35 fetuses in the second trimester were collected by Single Shot Fast Spin Echo (SSFSE) with pixel
size 0.74 mm-1.58 mm and inter-slice spacing 3 mm-4 mm. Each fetus had three scans in axial, sagittal and coronal views respectively, leading to 105 stacks in total. We randomly selected 72 stacks from 24 patients, 9 stacks from 3 patients and 24 stacks from 8 patients for training, validation and testing, respectively. Manual segmentations of these images were used as the ground truth. 

We implemented the CNNs in Pytorch on a Ubuntu desktop with an NVIDIA GTX 1080 Ti GPU and developed a PyQt GUI for user interactions. I-DRLSE was implemented in Python and it ran on the CPU. To train MG-Net, we used Dice loss~\cite{Milletari2016} and Adam optimizer with weight decay $10^{-5}$, mini-batch of 24 slices and learning rate $10^{-4}$. The training was ended when performance on the validation set stopped to increase for 5k iterations. The group number $N$ in MG-Net was  4 and the channel number parameter $C$ in Fig.~\ref{fig:network} was 16. For interactive refinement, $M'$ was set to $0.6M$ as refinement was not needed for more than half of the slices in the validation set. For I-DRLSE, the maximal evolution step number was 200 and $\alpha$=0.1, $\beta$=0.5, $\lambda$=0.3, $\mu$=0.005, $D$=4.0 based on grid search on the validation set. The segmentation accuracy was measured by Dice similarity and Average Symmetric Surface Distance (ASSD) between segmentation results and the ground truth. 

\subsubsection{Initial Segmentation and Uncertainty Estimation.} 
\begin{table*}
	\centering
	
	\caption{Quantitative evaluation of different uncertainty estimation methods for fetal brain segmentation before refinement. The values were measured at stack-level.}
	\label{tab:eval_1}
	\scriptsize
	\begin{tabular}
		{p{0.16\linewidth}|
			p{0.12\linewidth} p{0.12\linewidth}| p{0.14\linewidth} p{0.14\linewidth}| p{0.14\linewidth}p{0.12\linewidth}} 
		\hline
		\multirow{2}{*}{Method}&
		\multicolumn{2}{c|}{Efficiency} &
		\multicolumn{2}{c|}{Uncertainty Quality} & \multicolumn{2}{c}{Segmentation Quality}
		\\ 
		& Param(M)& Runtime(s) &  UEO(\%) & RVE(\%) & Dice(\%) &ASSD(mm)\\ \hline
		U-Net~\cite{Ronneberger2015}   & 11.51 & 0.31$\pm$0.09  &  -- & --  & 91.54$\pm$5.87 &  4.31$\pm$2.45
		\\
		MC Dropout~\cite{Gal2016} & 11.51  & 3.12$\pm$0.83 &  36.17$\pm$5.31 & 42.64$\pm$36.94  &  92.04$\pm$9.27 &  4.16$\pm$2.75   
		\\
		Ensemble~\cite{Lakshminarayanan2017} & 11.51  & 1.56$\pm$0.42 &  37.08$\pm$7.75 & 40.74$\pm$37.29 & \textbf{92.05$\pm$5.72}& 4.07$\pm$2.37 
		\\
		Bayesian Net~\cite{Jena2019} & 11.51 & 0.31$\pm$0.09 & 36.64$\pm$5.10 & 26.54$\pm$18.29  & 90.24$\pm$7.39 &   4.87$\pm$2.82  
		\\
		MG-Net& \textbf{4.86} &  \textbf{0.29$\pm$0.08} & \textbf{40.47$\pm$3.58} & \textbf{21.90$\pm$17.62} & 91.82$\pm$4.78 &  \textbf{4.07$\pm$2.13}   \\
		\hline
	\end{tabular}
\end{table*}
\begin{figure*}[t]
	\centering 
	\includegraphics[width=1.0\textwidth]{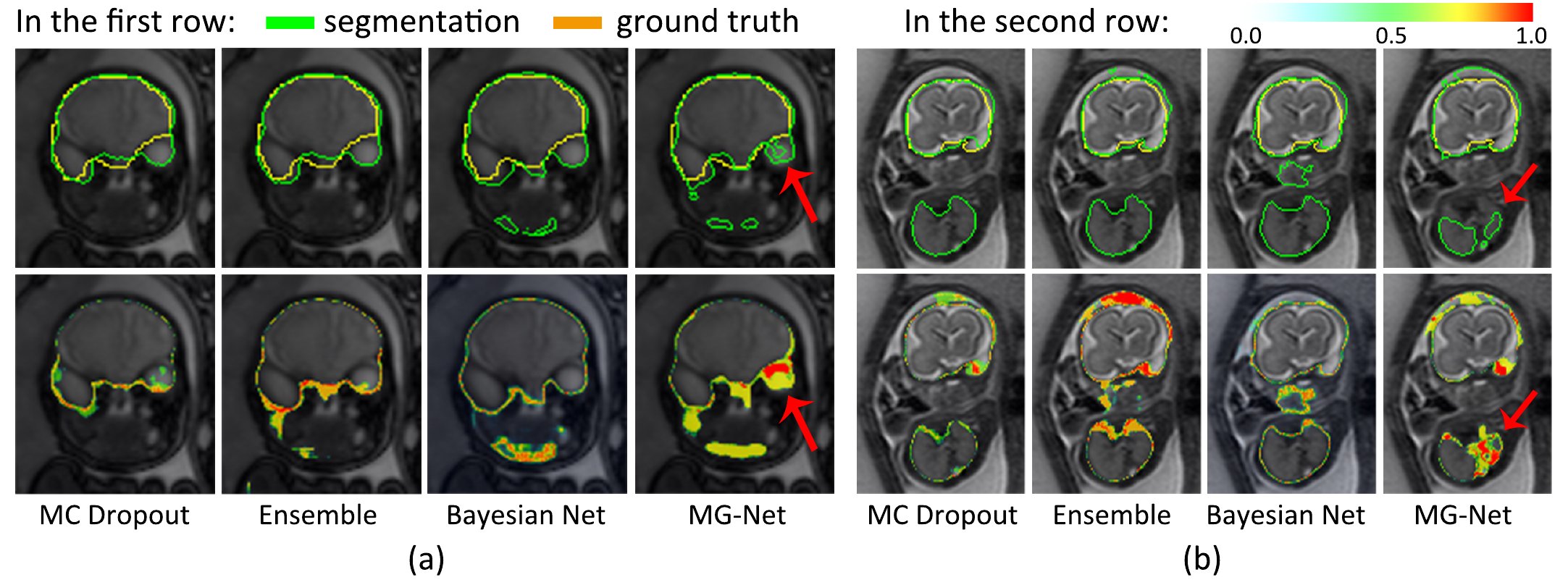}
	\caption{
		Comparison of different methods for initial segmentation (1st row) and uncertainty estimation (2nd row) that is normalized to [0, 1]. Red arrows highlight the consistency between mis-segmentation (eye ball in (a) and lung in (b)) and uncertainty obtained by MG-Net. }
	\label{fig:uncertainty}
\end{figure*}
Our MG-Net was compared with MC dropout~\cite{Gal2016} with 10 folds, ensemble~\cite{Lakshminarayanan2017} of 5 models, and a Bayesian network~\cite{Jena2019} for uncertainty estimation. We implemented all these methods using U-Net~\cite{Ronneberger2015} as the backbone. To measure ability of the uncertainty to indicate mis-segmentation, we used Uncertainty-Error Overlap (UEO, i.e., Dice as in~\cite{Jungo2019a}) and Relative Volume Error (RVE) between thresholded uncertain region and mis-segmented region. The optimal threshold value for each method was determined based on the validation set.  Quantitative evaluation results are shown in Table~\ref{tab:eval_1}. Compared with ensemble of U-Net, our MG-Net obtains comparable segmentation accuracy and higher uncertainty estimation quality. MG-Net takes 0.29s in average for simultaneous automatic segmentation and uncertainty estimation of a stack, which  is far more efficient than MC dropout and model ensemble and more suitable for interactive segmentation. The Bayesian network method also obtains uncertainty estimation in a fast speed, but with a reduced segmentation accuracy. Visual comparison in Fig.~\ref{fig:uncertainty} shows that  MG-Net obtains better consistency between mis-segmentation and uncertain regions than the other uncertainty estimation methods. 

\subsubsection{Refinement using Interactive Level Set.} 
\begin{figure*}[t]
	\centering 
	\includegraphics[width=1.0\textwidth]{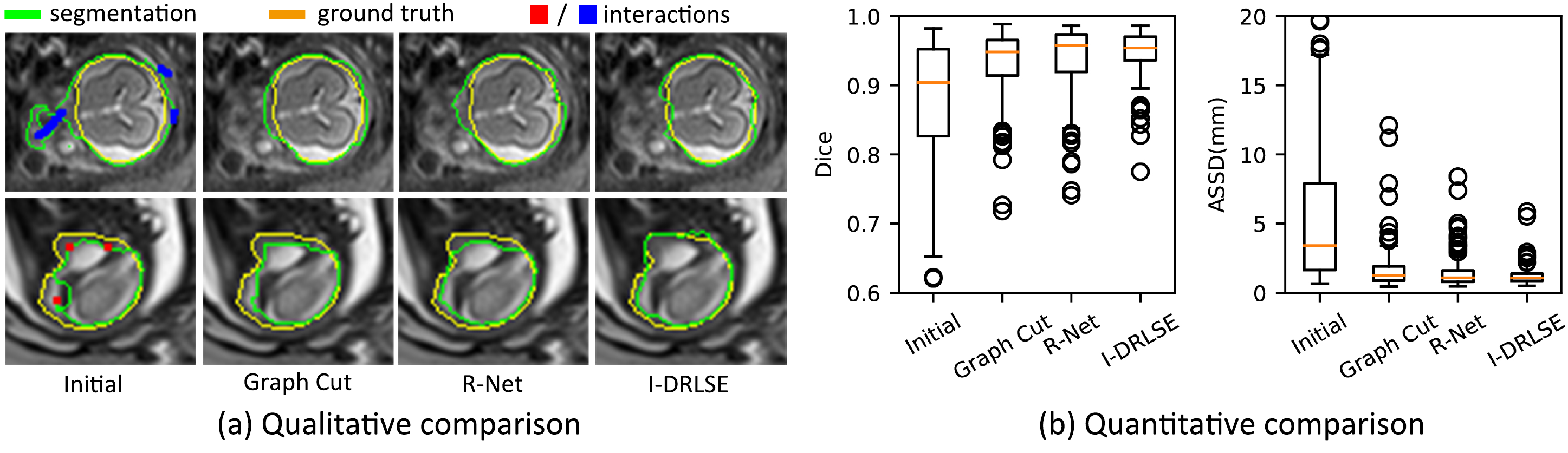}
	\caption{
		Slice-level qualitative and quantitative comparisons of different methods for refinement based on the same set of initial segmentation and user interactions.}
	\label{fig:refine}
\end{figure*}

Our I-DRLSE was firstly validated with slice-level refinement, and compared with: 1) CPU-based Graph Cut\footnote{\url{Code from the maxflow-v3.01 library: https://vision.cs.uwaterloo.ca/code/}} as implemented in~\cite{Wang2018}, and 2) training a refinement CNN (a.k.a., R-Net)~\cite{Wang2018c} that takes the initial segmentation and user interactions as input. We implemented the R-Net using U-Net~\cite{Ronneberger2015} as the backbone, and trained it with simulated interactions on initial segmentation obtained by MG-Net following~\cite{Wang2018c}. As not all the slices in a stack require refinement, we randomly selected 100 obviously mis-segmented slices from the test set, and used the same set of user  interactions on the initial segmentation for comparison. Fig.~\ref{fig:refine}(a) shows two cases where over- and under-segmentation exist in the initial segmentation respectively, which demonstrates that I-DRLSE obtains higher refinement accuracy than Graph Cut and R-Net with the same initial segmentation and interactions.  Fig.~\ref{fig:refine}(b) shows a quantitative comparison of these refinement methods. It can be observed that  I-DRLSE leads to higher Dice scores and lower ASSD values than Graph Cut and R-Net for refinement. 
The average slice-level  machine time for CPU-based I-DRLSE was 0.45s, which is slower than that of Graph Cut (0.07s) and R-Net (0.12s), but still acceptable for fast response of user interactions. The efficiency of I-DRLSE could be further improved by a GPU-optimized implementation in the future.
\subsubsection{Comparison of Different Interactive Frameworks.} For stack-level segmentation, our UGIR was compared with two variants: 1) UGIR(-U) denoting that the user manually searches mis-segmentation slice-by-slice to provide interactions (i.e., not guided by uncertainty) for refinement using I-DRLSE; 2) UGIR(*) that denotes using the naive slice-level uncertainty $\nu^*$ to guide user interactions. They were also compared with two existing interactive segmentation methods: using Graph Cut~\cite{Boykov2001} for interactive segmentation from scratch, and DeepIGeoS~\cite{Wang2018c} that uses two CNNs for initial segmentation and interactive refinement, respectively. We re-implemented DeepIGeoS following the training method in~\cite{Wang2018c}.  One user employed these methods to segment the fetal brain from the testing fetal MRI stacks respectively, where the interactions in a slice could be given multiple times until the result was accepted.  Quantitative evaluation results are shown in Table~\ref{tab:compare_frameworks}. Compared with DeepIGeoS~\cite{Wang2018c}, our UGIR obtained similar final accuracy ($p$-value $>$ 0.05 based on a paired t-test), but reduced the runtime from 75.38s to 48.46s. By using uncertainty-guided user interactions, UGIR improved the efficiency by near 30\% from UGIR(-U). UGIR(*) and UGIR took almost the same runtime, but UGIR achieved higher accuracy.
\begin{table*}
	\centering
	
	\caption{Quantitative comparison of different interactive methods for stack-level  segmentation of the fetal brain.}
	\label{tab:compare_frameworks}
	\begin{tabular}{c|c|c|c|c|c}
		\hline
		& Graph Cut~\cite{Boykov2001} & DeepIGeoS~\cite{Wang2018c} & UGIR(-U) &  UGIR(*) & UGIR \\ \hline
		Dice (\%)  & 93.17$\pm$3.15  &  \textbf{95.03$\pm$3.07} & 95.00$\pm$3.09  & 94.65$\pm$3.28 &  94.83$\pm$3.22
		\\
		ASSD (mm)& 2.84$\pm$1.18 & 2.72$\pm$1.74 & 2.73$\pm$1.12  &  2.75$\pm$1.17 &  \textbf{2.70$\pm$1.15}   
		\\
		Runtime (s) & 214.54$\pm$58.73 &  75.38$\pm$42.67 & 68.04$\pm$26.00 & \textbf{48.11$\pm$20.87}& 48.46$\pm$19.40
		\\	
		\hline
	\end{tabular}
\end{table*}
\section{Conclusion}
In this work, we propose a novel interactive segmentation framework using uncertainty to efficiently guide user interactions for refining results obtained by automatic CNNs. We introduce MG-Net based on grouped convolution to obtain multiple segmentation predictions simultaneously with real-time uncertainty estimation, which is used to suggest mis-segmented slices for user interactions, avoiding unnecessary manual check of well-segmented slices and leading to improved efficiency. A novel interactive level set I-DRLSE is also proposed to obtain refined results with spatial regularization. Experiments with fetal brain segmentation from stacks of motion-corrupted fetal MRI slices show that the proposed interactive framework achieved high accuracy with fast runtime, and the uncertainty information helped to improve the refinement efficiency by around 30\%.

\subsubsection{Acknowledgements.}\label{sec:acknowledgements}
This work was supported by the National Natural Science Foundation of China funding [81771921, 61901084], the Wellcome Trust [WT101957, 203148/Z/16/Z],  and the Engineering and Physical Sciences Research Council (EPSRC) [NS/A000027/1, NS/A000049/1]. TV is supported by a Medtronic / Royal Academy of Engineering Research Chair [RCSRF1819/7/34].
%
%
\bibliographystyle{splncs03}
\bibliography{./reference/level_set_seg}

\begin{thebibliography}{10}
\providecommand{\url}[1]{\texttt{#1}}
\providecommand{\urlprefix}{URL }

\bibitem{Boykov2001}
Boykov, Y.Y., Jolly, M.P.: {Interactive graph cuts for optimal boundary {\&}
  region segmentation of objects in N-D images}. In: ICCV. pp. 105--112 (2001)

\bibitem{Ebner2020}
Ebner, M., Wang, G., Li, W., Aertsen, M., Patel, P.A., Aughwane, R., Melbourne,
  A., Doel, T., Dymarkowski, S., {De Coppi}, P., David, A.L., Deprest, J.,
  Ourselin, S., Vercauteren, T.: {An automated framework for localization,
  segmentation and super-resolution reconstruction of fetal brain MRI}.
  Neuroimage  206,  116324 (2020)

\bibitem{Gal2016}
Gal, Y., Ghahramani, Z.: {Dropout as a Bayesian approximation: representing
  model uncertainty in deep learning}. In: ICML. pp. 1050--1059 (2016)

\bibitem{Jena2019}
Jena, R., Awate, S.P.: {A bayesian neural net to segment images with
  uncertainty estimates and good calibration}. In: IPMI. pp. 3--15 (2019)

\bibitem{Jungo2019a}
Jungo, A., Reyes, M.: {Assessing reliability and challenges of uncertainty
  estimations for medical image segmentation}. In: MICCAI. vol.~1, pp. 48--56
  (2019)

\bibitem{Keraudren2014}
Keraudren, K., Kuklisova-Murgasova, M., Kyriakopoulou, V., Malamateniou, C.,
  Rutherford, M.A., Kainz, B., Hajnal, J.V., Rueckert, D.: {Automated fetal
  brain segmentation from 2D MRI slices for motion correction}. Neuroimage
  101,  633--643 (2014)

\bibitem{Krizhevsky2012}
Krizhevsky, A., Sutskever, I., Hinton, G.E.: {ImageNet classification with deep
  convolutional neural networks}. In: NeurIPS. pp. 1097--1105 (2012)

\bibitem{Lakshminarayanan2017}
Lakshminarayanan, B., Pritzel, A., Blundell, C.: {Simple and scalable
  predictive uncertainty estimation using deep ensembles}. In: NeurIPS. pp.
  6405--6416 (2017)

\bibitem{ChunmingLi2010}
Li, C., Xu, C., Gui, C., Fox, M.D.: {Distance regularized level set evolution
  and its application to image segmentation}. IEEE Trans. Image Process.
  19(12),  3243--3254 (2010)

\bibitem{Milletari2016}
Milletari, F., Navab, N., Ahmadi, S.A.: {V-Net: Fully convolutional neural
  networks for volumetric medical image segmentation}. In: IC3DV. pp. 565--571
  (2016)

\bibitem{Ronneberger2015}
Ronneberger, O., Fischer, P., Brox, T.: {U-Net: Convolutional networks for
  biomedical image segmentation}. In: MICCAI. pp. 234--241 (2015)

\bibitem{Salehi2017isbi}
Salehi, S.S.M., Hashemi, S.R., Velasco-Annis, C., Ouaalam, A., Estroff, J.A.,
  Erdogmus, D., Warfield, S.K., Gholipour, A.: {Real-time automatic fetal brain
  extraction in fetal MRI by deep learning}. In: ISBI. pp. 720--724 (2018)

\bibitem{Top2011}
Top, A., Hamarneh, G., Abugharbieh, R.: {Active learning for interactive 3D
  image segmentation}. In: MICCAI. pp. 603--610 (2011)

\bibitem{Wang2019neurocomp}
Wang, G., Li, W., Aertsen, M., Deprest, J., Ourselin, S., Vercauteren, T.:
  {Aleatoric uncertainty estimation with test-time augmentation for medical
  image segmentation with convolutional neural networks}. Neurocomputing  338,
  34--45 (2019)

\bibitem{Wang2018}
Wang, G., Li, W., Zuluaga, M.A., Pratt, R., Patel, P.A., Aertsen, M., Doel, T.,
  David, A.L., Deprest, J., Ourselin, S., Vercauteren, T.: {Interactive medical
  image segmentation using deep learning with image-specific fine-tuning}. IEEE
  Trans. Med. Imaging  37(7),  1562--1573 (2018)

\bibitem{Wang2018c}
Wang, G., Zuluaga, M.A., Li, W., Pratt, R., Patel, P.A., Aertsen, M., Doel, T.,
  David, A.L., Deprest, J., Ourselin, S., Vercauteren, T.: {DeepIGeoS: A deep
  interactive geodesic framework for medical image segmentation}. IEEE Trans.
  Pattern Anal. Mach. Intell.  41(7),  1559--1572 (2019)

\bibitem{Zhao2013}
Zhao, F., Xie, X.: {An overview of interactive medical image segmentation}.
  Ann. BMVA  2013(7),  1--22 (2013)

\bibitem{Zhou2019}
Zhou, B., Chen, L., Wang, Z.: {Interactive deep editing framework for medical
  image segmentation}. In: MICCAI. pp. 329--337 (2019)

\end{thebibliography}

%
%

\end{document}